\documentclass[fleqn,usenatbib]{mnras}

\usepackage{newtxtext,newtxmath}

\usepackage[T1]{fontenc}

\DeclareRobustCommand{\VAN}[3]{#2}
\let\VANthebibliography\thebibliography
\def\thebibliography{\DeclareRobustCommand{\VAN}[3]{##3}\VANthebibliography}

\usepackage{graphicx}	
\usepackage{amsmath}	
\usepackage{float}
\usepackage{placeins}

\title[The descendants of $z\gtrsim10$ JWST galaxies ]{The descendants of $z\gtrsim10$ JWST galaxies in the COLIBRE simulations}

\author[Xu Zhao et al.]
{Xu Zhao$^{1,2,3}$\thanks{E-mail: xu.zhao@durham.ac.uk},
Carlos S. Frenk$^{2}$,
Andrew Pontzen$^{2}$,
Kyle A. Oman$^{2}$,
Evgenii Chaikin$^{2,4}$,
\newauthor
Isabel Santos-Santos$^{5}$,
Shengdong Lu$^{2}$,
Joop Schaye$^{4}$,
Alejandro Ben\'itez-Llambay$^{6}$,
Filip Hu\v{s}ko$^{4}$,
\newauthor
Alexander J. Richings$^{7,8}$,
Matthieu Schaller$^{4,9}$
\\
$^{1}$Key Laboratory for Computational Astrophysics, National Astronomical Observatories, Chinese Academy of Sciences, Beijing 100012, China\\
$^{2}$Institute for Computational Cosmology, Department of Physics, Durham University, South Road, Durham, DH1 3LE, UK\\
$^{3}$School of Astronomy and Space Science, University of Chinese Academy of Sciences, Beijing 100049, China\\
$^{4}$Leiden Observatory, Leiden University, PO Box 9513, 2300 RA Leiden, the Netherlands\\
$^{5}$Leibniz-Institut f{\"u}r Astrophysik Potsdam (AIP), An der Sternwarte 16, 14482 Potsdam, Germany\\
$^{6}$Dipartimento di Fisica G. Occhialini, Universit\`a degli Studi di Milano Bicocca, Piazza della Scienza, 3 I-20126 Milano MI, Italy\\
$^{7}$Centre for Data Science, Artificial Intelligence and Modelling, University of Hull, Cottingham Road, Hull, HU6 7RX, UK\\
$^{8}$E. A. Milne Centre for Astrophysics, University of Hull, Cottingham Road, Hull, HU6 7RX, UK\\
$^{9}$Lorentz Institute for Theoretical Physics, Leiden University, PO Box 9506, 2300 RA Leiden, the Netherlands
}

\date{Accepted XXX. Received YYY; in original form ZZZ}

\pubyear{\the\year{}}

\begin{document}
\label{firstpage}
\pagerange{\pageref{firstpage}--\pageref{lastpage}}
\maketitle

\begin{abstract}
Recent observations with \textit{JWST} have revealed a population of UV-bright galaxies at $z\gtrsim 10$. This discovery naturally raises the question: what do such early galaxies evolve into by the present day? In this work, we address this descendant question using the new-generation COLIBRE cosmological hydrodynamical simulations to trace bright galaxies selected at $z=10$ and follow their descendants to the present day. Most of the high-redshift galaxies do not survive as distinct, self-bound objects to $z=0$; instead, the majority are incorporated into more massive systems through merging or disruption. The surviving descendants span a broad range of present-day stellar masses, although they are most commonly intermediate- to high-mass, $M_\star\sim10^{10}$--$10^{11}\,{\rm M_\odot}$. They typically reside in galaxy groups and clusters, with host halo masses, $M_{200{\rm c}}\sim10^{13}$--$10^{14}\,{\rm M_\odot}$. The large scatter in descendant stellar mass shows that present-day outcomes retain only a weak memory of the stellar mass of the high-redshift progenitor. We show that the evolution of descendant host halo masses is consistent with the forward conditional distribution predicted by extended Press--Schechter (EPS) theory, both in the median growth and in the large scatter in descendant mass. In particular, EPS confirms that massive present-day galaxies typically do not originate from the most massive objects at high redshift. A galaxy observed at $z\gtrsim10$ therefore cannot be interpreted as the direct progenitor of a single class of $z=0$ galaxies.

\end{abstract}




\section{Introduction}
Recent observations with \textit{JWST} have dramatically extended our view of the high-redshift Universe, revealing a population of spectroscopically confirmed galaxies and photometrically selected candidates at \(z\gtrsim10\) with high UV luminosities and, in some cases, large inferred stellar masses and number densities \citep[e.g.][]{Naidu2022JWST,Castellano2022JWST,Atek2023JWST,Finkelstein2023JWST,Carniani2024JWST,Naidu2026JWST}. These discoveries have been interpreted as a potential stress test of galaxy formation theory in the standard $\Lambda$ cold dark matter ($\Lambda$CDM) cosmology \citep[e.g.][]{Menci2022Highz,BoylanKolchin2023JWST}. More broadly, these observations have renewed interest in the efficiency, diversity, and physical pathways of galaxy formation in the first few hundred million years of cosmic history.

Cosmological hydrodynamical simulations of galaxy formation are a key tool for interpreting these observations by modelling galaxy formation within a $\Lambda$CDM framework, following the baryonic processes that connect dark matter halo growth to observable galaxy properties. Large representative-volume simulations such as Illustris \citep{Vogelsberger2014Illustris}, EAGLE \citep{Schaye2015EAGLE}, IllustrisTNG \citep{Springel2018TNG}, SIMBA \citep{Dave2019SIMBA} and FLAMINGO \citep{Schaye2023FLAMINGO} have reproduced many aspects of the galaxy population and have been used to make predictions for galaxy properties across cosmic time. The new-generation COLIBRE simulations \citep{Schaye2026ColibreIntro, Chaikin2026calibration} introduce a number of improvements in the modelling, including an explicit treatment of the multiphase interstellar medium, a self-consistent model of dust formation and evolution, supersampled dark matter and more realistic treatments of stellar and active galactic nuclei (AGN) feedback. Recent COLIBRE results show that the inferred evolution of the galaxy stellar mass function and star formation rates from $z\approx12$ to $z=0$ are broadly reproduced \citep{Chaikin2026COLIBRESMF}, while the dust-attenuated UV luminosity functions remain systematically below the observed bright end at $z=7$--15 \citep{Lu2026COLIBREuvHighz}, a discrepancy can be ascribed to the assumption of a universal, \cite{Chabrier2003IMF} initial stellar mass function \citep[IMF;][]{Lushengdong2025GALFORM,Lu2026COLIBREuvHighz,Anna2026COLIBRETopHeavy}. Most recently, \citet{Chaikin2026progenitorHighz} show that the most massive JWST galaxies out to redshift $\sim 14.5$ all have COLIBRE counterparts with similar masses, star formation rates, ages and gas-phase metallicities. 

Contrary to initial claims \citep[e.g.][]{Finkelstein2023JWST,BoylanKolchin2023JWST}, several recent studies have shown that the observed abundance of UV-bright galaxies at very high redshift can be accommodated within a $\Lambda$CDM framework, although the required physical ingredients differ between models. A particularly simple explanation, proposed before the launch of \textit{JWST}, is that the IMF becomes top-heavy in the exceptional physical conditions that prevail in starbursts \citep{Cowley2018GALFORM}; this possibility has recently been explored in a more self-consistent way within the context of COLIBRE using a density-dependent IMF \citep{Anna2026COLIBRETopHeavy}.

\citet{Mason2023CosmicDawn} argued that the brightest $z\gtrsim10$ galaxies may represent the youngest and most rapidly star-forming galaxies in the tail of the halo population, while \citet{Ferrara2023SuperEarly} showed that reduced dust attenuation at $z\gtrsim11$ can offset the declining abundance of massive haloes and reproduce the bright end of the UV luminosity function at $z=10$--14. \cite{Dekel2023} postulated that feedback-free starbursts could explain the bright high-redshift population. \citet{Somerville2025DMSFE} showed that a density-modulated star formation efficiency, together with decreasing dust attenuation and increasingly bursty star formation, can reproduce the observed abundance of UV-luminous galaxies at $z\approx6$--17. A limitation of these studies is that they do not demonstrate that the \textit{ansatz} specifically invoked to explain the high-redshift galaxy population would lead to a realistic galaxy population at lower redshifts, including the present day. The realism of the proposed mechanisms is therefore difficult to assess.

By contrast, the pre-\textit{JWST} GALFORM model of \cite{Cowley2018GALFORM} produces a galaxy population that matches not only the UV luminosity functions at $z>10$ but also a plethora of other observables across time, including luminosity functions in various passbands and scaling relations \citep{Lacey2016GALFORM,Lushengdong2025GALFORM}. At high redshift, the only modification required is to the default dust obscuration model, by including evolution of the dust mass to take into account the short age of the universe at $z>12$. 

Following the discovery of galaxies barely 250 million years after the Big Bang, it is natural to ask about their fate, in particular about their descendants at the present day. In our hierarchical clustering Universe, galaxies follow diverse assembly histories determined by mergers, tidal interactions and environmental effects \citep{Frenk1985,Frenk1988,White-Frenk1991,Kauffmann1993,Cole2000}. 

Related questions regarding the descendants of rare early structures, first-star remnants and quasars have been addressed in the past. \citet{WhiteSpringel2000FirstStars} showed, from a resimulation of a massive galaxy cluster, that the oldest stars are to be found today in the central regions of rich clusters. \citet{Trenti2008FirstStar}, using cosmological simulations and EPS calculations, argued that the remnants of Population~III stars should typically reside in group-scale haloes today, while the descendants of bright \(z\gtrsim6\) quasars should today lie primarily in intermediate-mass clusters rather than exclusively in the richest clusters. \citet{Fanidakis2013quasars} concluded, using the GALFORM semi-analytic model, that the most luminous quasars do not occupy the most massive dark matter haloes at any redshift and that the descendants of luminous \(z\sim6\) quasars reside in group- to cluster-scale haloes by \(z=0\). Therefore, the expected properties of descendants are not obvious.

More recently, descendant studies have been extended to high-redshift galaxy populations. From the ELUCID constrained-realization simulation, \citet{ChenYangyao2023JWSTDesc} found that a substantial fraction of the descendants of bright galaxies at \(z=7\)--10 reside in present-day host haloes of \(M_{\rm halo}>10^{13}h^{-1}{\rm M_\odot}\). They predicted that the central galaxies of massive clusters today contain old stellar populations formed in those early systems. In the GALFORM semi-analytic model, \citet{Lushengdong2025GALFORM} traced UV-bright galaxies selected at \(z=14\) to the present day, finding a narrow range of descendant halo mass, with a median of \(2.5\times10^{13}h^{-1}{\rm M_\odot}\), a broad distribution of descendant stellar masses, and a descendant population in which half are satellite galaxies. A recent study with TNG300 \citep{Baxter2026desc} similarly found that the most massive NIRCam-selected galaxies at \(z=7\), 4, and 2 rarely evolve into the most massive galaxies at \(z=0\).

These studies provide important context for interpreting the link between the early \textit{JWST} galaxies and their present-day descendants. Here, we analyse COLIBRE \citep{Schaye2026ColibreIntro,Chaikin2026calibration}, a set of state-of-the-art cosmological hydrodynamical simulations with particle masses of $\sim 10^5, \sim 10^6\ {\rm and} \sim 10^7\,{\rm M_\odot}$ in cubic volumes of \(100\), \(200\), and \(400\,{\rm cMpc}\) on a side, respectively. Our fiducial analysis is based on the intermediate-resolution L200m6 simulation, a \((200\,{\rm cMpc})^3\) volume followed to \(z=0\) with \(m_{\rm gas}=1.8\times10^6\,{\rm M_\odot}\) and \(m_{\rm dm}=2.4\times10^6\,{\rm M_\odot}\). This volume and resolution allow us to sample rare \(z\gtrsim10\) galaxies while achieving a minimum resolved stellar mass of \(M_\star\simeq9\times10^6\,{\rm M_\odot}\). Compared with earlier representative-volume ($100\,{\rm cMpc})^3$\ simulations run to \(z=0\), such as EAGLE and IllustrisTNG, L200m6 has comparable baryonic mass resolution but about eight times as many baryonic resolution elements and 32 times as many dark matter particles.

In this paper we ask which present-day populations the galaxies observed at \(z\gtrsim10\) evolve into. We track galaxies selected at \(z=10\) and \(z=12\) in COLIBRE, using both a stellar mass-selected parent sample and a dust-attenuated UV-bright subset. We quantify their fate, distinguishing systems that survive as centrals or satellites from those that lose their independent identity through merging or disruption. We then examine their descendant stellar and host halo masses, and determine whether UV-bright galaxies follow distinct evolutionary pathways from the broader mass-selected population. Finally, by combining the simulated descendant tracks with EPS theory, we interpret the broad scatter in descendant halo masses as a natural consequence of stochastic hierarchical assembly.

This paper is organised as follows. In Section~\ref{sec:methods}, we describe the simulations, sample selection, and descendant catalogues. In Section~\ref{sec:z0}, we present the descendant fates, stellar masses, host halo masses, and the comparison between UV-bright galaxies and the parent population. In Section~\ref{sec:evolution}, we interpret the results in the context of hierarchical structure formation and EPS theory. Finally, we summarise our conclusions in Section~\ref{sec:conclusions}.

\section{Methods}
\label{sec:methods}
\subsection{The COLIBRE simulations}
\label{sec:colibre}
The COLIBRE (COLd ISM and Better REsolution, \citealt{Schaye2026ColibreIntro, Chaikin2026calibration})\footnote{\url{www.colibre-simulations.org}} cosmological hydrodynamical simulations were performed with the \textsc{swift} code \citep{Schaller2024SwiftCode}, which employs task-based parallelism to solve for the coupled evolution of dark matter, gas, stars, and black holes within the $\Lambda$CDM cosmology. Gas hydrodynamics is modelled using the density--energy smoothed particle hydrodynamics scheme \textsc{Sphenix} \citep{Borrow2022SPHENIX}.

A detailed description of the COLIBRE galaxy formation model is presented in \citet{Schaye2026ColibreIntro}; here we provide only a brief summary of the ingredients most relevant for this work. The simulations include subgrid models for gas cooling and heating \citep{Ploeckinger2025HYBRID-CHIMES}, star formation \citep{Nobels2024starformation}, stellar evolution and chemical enrichment \citep{Correa2026chemical}, supernova and early stellar feedback \citep{Chaikin2023SNfeedback,BenitezLlambay2026earlyfeedback}, black-hole repositioning and merging \citep{Bahe2022BH}, and, in a subset of runs, kinetic jet AGN feedback \citep{husko2026AGNfeedback}. The strengths of the supernova and AGN feedback were calibrated so that the simulation reproduces the observationally inferred galaxy stellar mass function and the galaxy size--stellar mass relation at $z=0$ \citep{Chaikin2026calibration}. A distinctive feature of COLIBRE is the explicit modelling of a multiphase interstellar medium, including a cold dense component, together with the evolution of dust grains \citep{Trayford2026dust}. This makes the simulations well suited for predicting dust attenuation and rest-frame ultraviolet emission in a physically motivated way. 

Throughout this work, we adopt the cosmological parameters used in the COLIBRE simulations (taken from \citealt{Abbott2022DESI_Parameters}): present-day matter density, $\Omega_{\rm m,0}=0.306$, present-day baryon density, $\Omega_{\rm b,0}=0.0486$, linear-theory amplitude of matter fluctuations on $8\,h^{-1}{\rm Mpc}$ scales, $\sigma_8=0.807$, and dimensionless Hubble parameter $h=0.681$.

Our primary analysis is based on the intermediate-resolution COLIBRE L200m6 simulation, which follows a periodic cube of side length $200\,{\rm cMpc}$. The mass resolution corresponds to gas particle masses of $m_{\rm gas}\approx1.8\times10^{6}\,{\rm M_\odot}$ and dark matter particle masses of $m_{\rm DM}\approx2.4\times10^{6}\,{\rm M_\odot}$. The gravitational softening length is the same for baryons and dark matter, $\epsilon=\min(0.7\,{\rm pkpc},1.8\,{\rm ckpc})$ at m6 resolution. L200m6 provides a suitable compromise between cosmological volume, needed to sample rare high-redshift galaxies, and mass resolution, needed to follow their subsequent evolution. We also use neighbouring COLIBRE resolution levels for convergence tests in Appendix~\ref{sec:appendix fateconvergence}: higher-resolution m5 simulations, with $m_{\rm gas}\approx2.3\times10^{5}\,{\rm M_\odot}$, and the lower-resolution L400m7 simulation, with $m_{\rm gas}\approx1.47\times10^{7}\,{\rm M_\odot}$, of a larger $400\,{\rm cMpc}$ volume.

Dark matter haloes in the simulation are first identified with a friends-of-friends algorithm \citep{Davis1985FOF}. Self-bound substructures and their merger trees are then constructed with the \textsc{HBT-HERONS} algorithm \citep{Forouhar2025HBTHerons}\footnote{\url{https://HBT-HERONS.strw.leidenuniv.nl/}}, an updated implementation of the hierarchical bound-tracing (HBT+) method \citep{Jiaxin2012HBT,Jiaxin2018HBT}. Unlike subhalo finders that identify structures independently in each snapshot, \textsc{HBT-HERONS} follows subhaloes through time by tracking the particles associated with each structure. This history-based approach reduces reliance on the instantaneous particle distribution alone, which can make satellites difficult to separate from the dense background of their host haloes, particularly near pericentric passage or after substantial tidal stripping.

Here, ``self-bound'' refers to particles that remain gravitationally bound to a candidate subhalo, after subtracting the bulk motion of that subhalo. In \textsc{HBT-HERONS}, this is determined through an iterative unbinding procedure: starting from a candidate or source subhalo, the algorithm identifies the particles gravitationally bound to it, removes unbound particles, recomputes the potential of the remaining particles, and repeats this process until the bound set converges. The resulting object is the self-bound remnant associated with that track; this requirement helps remove spurious candidate subhaloes produced by discreteness noise. Operationally, a subhalo is considered resolved only if its self-bound remnant contains at least 20 bound particles in total and at least 10 tracer particles, where the tracers are the most bound particles of dark matter or stellar type used to follow the subhalo between outputs. If a previously resolved subhalo falls below either of these thresholds, it is no longer treated as an independently resolved object and is classified as disrupted, unless a merger with another resolved subhalo has already been identified.

Galaxy properties are measured using the \textsc{soap} catalogue pipeline \citep{McGibbon2025SOAP}, which computes stellar masses, star formation rates, gas properties, and photometric and structural quantities (e.g. $u$- and $r$-band magnitudes and bulge-to-total ratios). Throughout this work, we adopt a spherical aperture of radius $50\,{\rm pkpc}$ centred on the most bound particle of each HBT-HERONS subhalo, including only particles gravitationally bound to that subhalo. This prevents double-counting of particles belonging to neighbouring galaxies, even when their apertures overlap.

\subsection{Dust-attenuated UV magnitudes from SKIRT}
\label{sec:skirt}

To connect simulated galaxies with UV-selected observations at high redshift, we compute dust-attenuated rest-frame ultraviolet magnitudes using the calibration-free COLIBRE--SKIRT pipeline introduced by \citet{Gebek2026COLIBRE_SKIRT}. This pipeline has been used to predict $z=0$ luminosity functions in COLIBRE from ultraviolet to sub-millimetre wavelengths \citep{2026LuCOLIBRELFz0} and ultraviolet luminosity functions at redshifts 7-20 in COLIBRE \citep{Lu2026COLIBREuvHighz}. It applies the SKIRT Monte Carlo radiative transfer code \citep{Camps2020Skirt} to COLIBRE galaxy outputs, following the absorption and scattering of stellar radiation by dust.

The calculation is based directly on the simulation particle data. Stellar particles older than $10\,{\rm Myr}$ are treated as evolved stellar populations and assigned BPASS \citep{Eldridge2017BPASS,Stanway2018BPASS} simple stellar population spectra according to their ages, metallicities and initial stellar masses. To reduce sampling noise, younger stellar populations are resampled into star-forming regions using star-forming gas and the parent gas of young star particles. Their spectra are computed with the dust-free TODDLERS \citep{Kapoor2023TODDLERS, Kapoor2026TODDLERS} library, which models the emission from young stars and their surrounding H\,{\sc ii} regions. No additional subgrid birth cloud attenuation is applied; attenuation around young stars is instead computed from the COLIBRE dust distribution.

The dust distribution and grain properties are also taken from COLIBRE, which follows the formation, growth, and destruction of dust grains self-consistently during the simulation. In the radiative transfer post-processing, dust associated with gas particles is mapped onto an adaptive grid. The dust masses and composition are inherited from COLIBRE; SKIRT represents the grains using a mixture of small and large carbonaceous and silicate grains, with wavelength-dependent absorption and scattering properties. Photon packets are propagated through this dusty medium, accounting for absorption and multiple scattering.

In COLIBRE, the dust-to-metal ratio increases with metallicity, reaching approximately 0.3 in metal-rich galaxies, while silicates dominate the dust mass and large grains dominate at $z\gtrsim5$ \citep{Vijayan2026dust}. Unlike models assuming a fixed dust-to-metal ratio, the COLIBRE--SKIRT pipeline \citep{Gebek2026COLIBRE_SKIRT} uses the dust masses and spatial distributions predicted by the simulation. The \citet{Draine2007Dust} model provides the wavelength-dependent optical properties and grain-size distributions of carbonaceous and silicate dust, which are divided into small and large components and rescaled to match the simulated dust masses. SKIRT then propagates stellar radiation through the resulting three-dimensional dust distribution, accounting for absorption and scattering to calculate dust-attenuated UV luminosities. UV attenuation generally increases with stellar mass, reaching $A_{\rm UV}\sim2$--$3$ mag for galaxies with $M_\star\sim10^9\,{\rm M_\odot}$ at $z=10$ \citep{Lu2026COLIBREuvHighz}, although dust-mass upper limits at $z\approx12$ suggest that grain growth in COLIBRE may be overly efficient \citep{Vijayan2026dust}.

The emergent spectral energy distribution is recorded using an ideal mock detector placed along the simulation $z$-axis, corresponding to a random viewing orientation for each galaxy rather than an inclination-averaged UV magnitude. We derive the dust-attenuated rest-frame UV magnitude, $M_{\rm UV,attn}$, from this line of sight. This approach captures the coupled effects of recent star formation, dust geometry, and viewing direction, which are particularly important for irregular and clumpy high-redshift galaxies.

\subsection{High-redshift mass-selected fiducial sample and UV-bright subset}
\label{sec:highz_samples}

\begin{figure}
\includegraphics[width=\columnwidth]{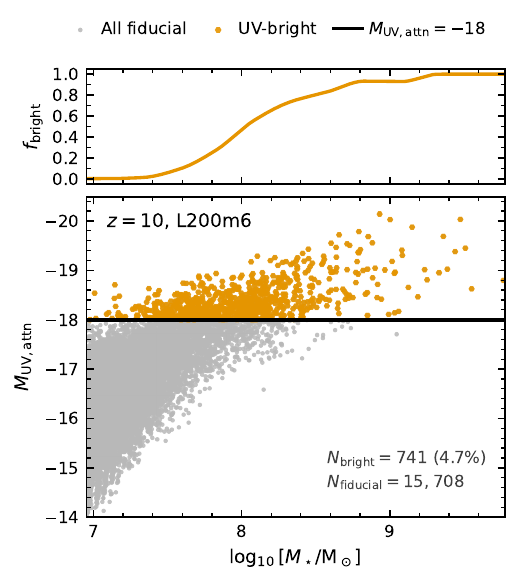}
\caption{Dust-attenuated rest-frame UV properties of galaxies at $z=10$ in the COLIBRE L200m6 simulation. The lower panel shows $M_{\rm UV,attn}$ as a function of stellar mass. Grey points represent the full mass-selected sample of galaxies containing at least five star particles, corresponding to $M_\star>9\times10^6\,{\rm M_\odot}$, while orange points highlight the UV-bright subset. The horizontal black line marks the UV-bright selection threshold, $M_{\rm UV,attn}=-18$. The upper panel shows the fraction of galaxies satisfying this criterion, $f_{\rm bright}$, as a function of stellar mass. This fraction increases strongly with $M_\star$, although there is substantial scatter in $M_{\rm UV,attn}$ at fixed stellar mass.}

  \label{fig:sample_selection_z10}
\end{figure}

We define our fiducial high-redshift sample at $z=10$ in the COLIBRE L200m6 simulation using a stellar mass resolution cut. Specifically, we require galaxies to contain at least five star particles, corresponding to $M_\star>9\times10^6\,{\rm M_\odot}$. This fiducial mass-selected sample is therefore the complete set of galaxies above our adopted stellar particle limit. For every galaxy satisfying this cut, we compute dust-attenuated rest-frame UV magnitudes using SKIRT.

This threshold removes the least resolved progenitors, whose subsequent fate is most susceptible to numerical resolution. It is close to the $m_{\star,\rm peak}>10^{7}\,{\rm M_\odot}$ threshold considered by \citet{He2026COLIBREtidal} in their COLIBRE study of satellite tidal evolution. Their fig.~13 shows that above this mass scale, the stellar tidal tracks of satellites are reasonably well converged between the m6 and m5 resolutions, although lower-resolution m7 systems show larger deviations. This provides an independent check that the m6 resolution is adequate for calculating the population-level survival and disruption statistics relevant to our analysis. Our convergence tests support this choice (Appendix~\ref{sec:appendix fateconvergence}).

Fig.~\ref{fig:sample_selection_z10} illustrates the selection of the UV-bright sample. The lower panel shows the dust-attenuated rest-frame UV magnitude, $M_{\rm UV,attn}$, as a function of stellar mass for galaxies in the fiducial mass-selected sample. Grey points show all galaxies above the five-star-particle resolution limit, while orange points highlight those satisfying $M_{\rm UV,attn}<-18$. The upper panel shows the fraction of UV-bright galaxies, $f_{\rm bright}$, as a function of stellar mass. This fraction increases strongly with stellar mass and is very small near the resolution limit. At $z=10$, $741$ of the $15,708$ mass-selected galaxies satisfy the UV-luminosity cut, corresponding to $4.7$~per~cent. The low value of $f_{\rm bright}$ at the low-mass end suggests that unresolved galaxies below the five-star-particle limit are unlikely to contribute appreciably to the UV-bright sample.

The two samples are designed to answer complementary questions. The fiducial mass-selected sample provides a stellar mass-selected baseline for all sufficiently resolved high-redshift galaxies in COLIBRE, appropriate for studying how the descendants of massive high-redshift galaxies evolve. This choice is motivated by previous work showing that COLIBRE broadly reproduces the observationally inferred high-redshift galaxy stellar mass function up to $z\approx 12$ \citep{Chaikin2026COLIBRESMF}. The UV-bright subset, selected with $M_{\rm UV,attn}<-18$, instead tests whether focusing on the most UV-luminous galaxies changes the inferred descendant population.

This distinction is important because dust-attenuated UV luminosity is not determined by stellar mass alone. As shown in Fig.~\ref{fig:sample_selection_z10}, there is substantial scatter in $M_{\rm UV,attn}$ at fixed $M_\star$, reflecting variations in recent star formation, dust attenuation, geometry and viewing direction. Moreover, the dust-attenuated COLIBRE UV luminosity function underpredicts the observed bright end at $z>7$ \citep{Lu2026COLIBREuvHighz}. We therefore interpret the UV-bright subset as a controlled selection of the brightest resolved galaxies within COLIBRE, rather than as an abundance-matched mock \textit{JWST} sample.

We apply the same stellar mass and UV-luminosity cuts at $z=12$ as a consistency check. At this redshift, the fiducial mass-selected sample contains $N=2,958$ galaxies, of which $N=102$ satisfy $M_{\rm UV,attn}<-18$, corresponding to $3.4$~per~cent. Since the qualitative descendant trends are similar to those at $z=10$, but the sample is smaller, we present the $z=12$ descendant statistics in Appendix~\ref{sec:appendix z12 results} and focus in the main text on the better sampled $z=10$ population.

\subsection{Descendant tracking and fate classification}
\label{sec:Descendant tracking}
\begin{figure*}
    
	\includegraphics[width=\textwidth]
    {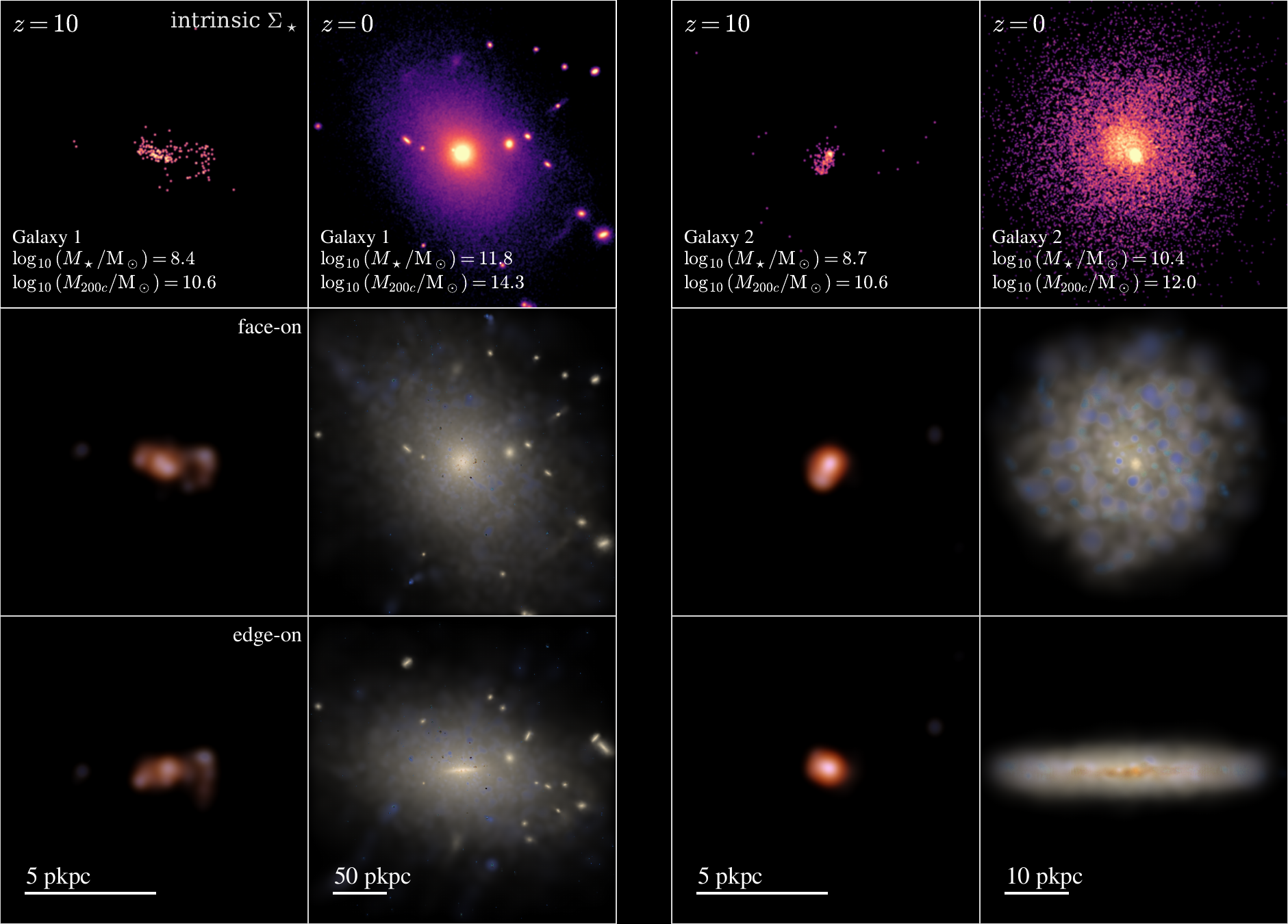}

\caption{
Visualisations of two galaxies selected at $z=10$ and their descendants at $z=0$. Odd-numbered columns show Galaxy 1 (TrackID 41064) and Galaxy 2 (TrackID 6535) at $z=10$, while even-numbered columns show their corresponding descendants at $z=0$. The top row shows intrinsic stellar surface-density maps, while the middle and bottom rows show dust-attenuated mock RGB images in face-on and edge-on projections, respectively.
The mock RGB images were generated following the procedure described in Section~\ref{sec:skirt}. The mock RGB images are noise-free and assume no specific exposure time or detection limit.
The $z=10$ panels show observer-frame JWST/NIRCam RGB composites constructed by redshifting source spectra to $z=10$ and applying the F444W, F356W, and F277W filters, assigned to the red, green, and blue channels, respectively. Each $z=10$ image covers $12\times12\,{\rm pkpc}$ in projection and integrates through a line-of-sight depth of $12\,{\rm pkpc}$. Before constructing the RGB composites, each filter image was convolved with a circular Gaussian approximation to the NIRCam PSF, adopting FWHM values of $0.140$, $0.114$, and $0.088$ arcsec, respectively.
The $z=0$ panels show HST/ACS RGB composites constructed from the F775W, F625W, and F475W filters, covering $300\times300\,{\rm pkpc}$ for Galaxy 1 and $50\times50\,{\rm pkpc}$ for Galaxy 2. The images are displayed using a stretch with dynamic range 500, corresponding to $2.5\log_{10}(500)\simeq6.75\,{\rm mag}$. Scale bars indicate physical distances and differ by a factor of five between the two systems.
Although both systems appear compact and clumpy at $z=10$, they evolve into very different present-day objects: Galaxy 1 becomes the central galaxy of a cluster, whereas Galaxy 2 becomes a lower-mass, disc-like galaxy in a Milky Way-mass halo.
}

    \label{fig:mock_images}
    
\end{figure*}
Using HBT-HERONS merger trees, we follow each galaxy selected at high redshift through its persistent \texttt{TrackID} for as long as it remains a resolved, self-bound subhalo. We then classify the final fate of each system at $z=0$ into three mutually exclusive categories: \textit{centrals}, \textit{satellites}, and \textit{merged/disrupted} systems.

The first two categories consist of galaxies that survive as resolved, self-bound systems at $z=0$. Among these survivors, \textit{centrals} are the primary galaxies of their host haloes, whereas \textit{satellites} are galaxies orbiting within larger host systems.

The third category consists of systems that no longer survive as independent, resolved subhaloes by $z=0$. This includes both \textit{merged} galaxies, whose original subhalo has coalesced with another resolved subhalo and is no longer identified as a distinct object, and \textit{disrupted} galaxies, whose subhaloes fall below the resolution threshold before a resolved merger is identified. Here, we group both merged and disrupted systems into a single \textit{merged/disrupted} category because in both cases the original high-redshift galaxy has lost its independent resolved identity.

We assess the robustness of this classification in Appendix~\ref{sec:appendix fateconvergence} by comparing the three descendant fate fractions across COLIBRE simulations with different mass resolution and volume. This test confirms that the descendant tracking is robust, with fate fractions differing by less than 5~per~cent between simulations.

Before turning to the population-level statistics in the next section, we first illustrate the progenitor--descendant connection with two example galaxies. Fig.~\ref{fig:mock_images} follows these systems from $z=10$ to the present day. The top row shows intrinsic stellar surface-density maps, while the middle and bottom rows show dust-attenuated mock RGB images in face-on and edge-on projections. The mock RGB images were generated using \textsc{SKIRT}, adopting the stellar emission prescription described in Section~\ref{sec:skirt}. Stellar populations older than 10 Myr are assigned BPASS SSP spectra, while emission from young star-forming regions is modelled using dust-free TODDLERS templates, including nebular emission lines. Further details of the image construction are given in the figure caption.

Both selected galaxies appear compact and clumpy at $z=10$, but they evolve into very different present-day objects. Galaxy 1 becomes the central galaxy of a cluster, of $M_\star \simeq 10^{11.8}\,{\rm M_\odot}$ and $M_{200{\rm c}} \simeq 10^{14.3}\,{\rm M_\odot}$ at $z=0$, and appears as an extended, spheroid-dominated system. By contrast, Galaxy 2 becomes a lower-mass, disc-like galaxy in a Milky Way-mass halo, of $M_\star \simeq 10^{10.4}\,{\rm M_\odot}$ and $M_{200{\rm c}} \simeq 10^{12.0}\,{\rm M_\odot}$; its disc morphology is especially clear in the edge-on projection. Thus, galaxies with similar compact high-redshift appearances can have very different present-day stellar masses, halo environments and morphologies. 

\section{The fate and present-day properties of high-redshift galaxies}
\label{sec:z0}

\subsection{What happens: the fate of high-redshift galaxies}
\label{sec:fate}
\begin{figure}
\includegraphics[width=\columnwidth]{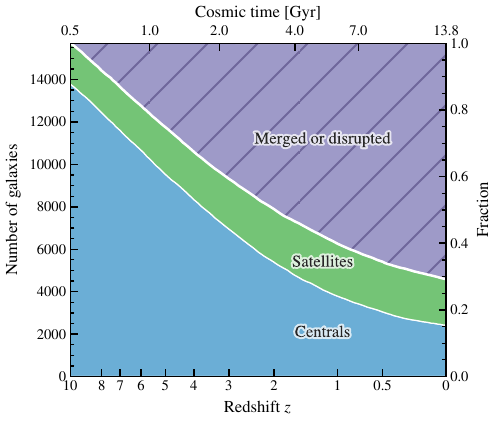}
\caption{
Evolution of the descendant fate fractions of galaxies selected at $z=10$. The coloured regions show the number of systems classified as surviving centrals, surviving satellites, or non-surviving systems that have either merged or become disrupted. The hatched region indicates the merged/disrupted population, with the thick white curve separating  surviving from non-surviving descendants. Most high-redshift galaxies do not survive as distinct systems to $z=0$.
}
\label{fig:fate_evolution}
\end{figure}

Fig.~\ref{fig:fate_evolution} shows the evolution of the descendant fate fractions for the $z=10$ mass-selected sample. 

Starting from $z=10$, the descendant population shows a clear monotonic evolution: the fraction of surviving systems steadily decreases with cosmic time, while the merged/disrupted fraction increases. At $z=10$, most galaxies ($\sim 90$~per~cent) are isolated centrals, but as structure formation proceeds, an increasing fraction is incorporated into larger systems. By $z=0$, only $\sim 30$~per~cent of the original population remains as resolved, self-bound galaxies, split between centrals ($\sim 15$~per~cent) and satellites ($\sim 15$~per~cent), while the majority has either merged or become disrupted. Thus, most high-redshift galaxies in our sample do not survive as independent systems, but are instead incorporated into larger structures through hierarchical assembly.

The $z=12$ sample shows qualitatively similar behaviour, but with increased stochasticity due to our smaller sample and more extreme properties (see Appendix~\ref{sec:appendix z12 results}).

\begin{figure*}
\includegraphics[width=\textwidth]
{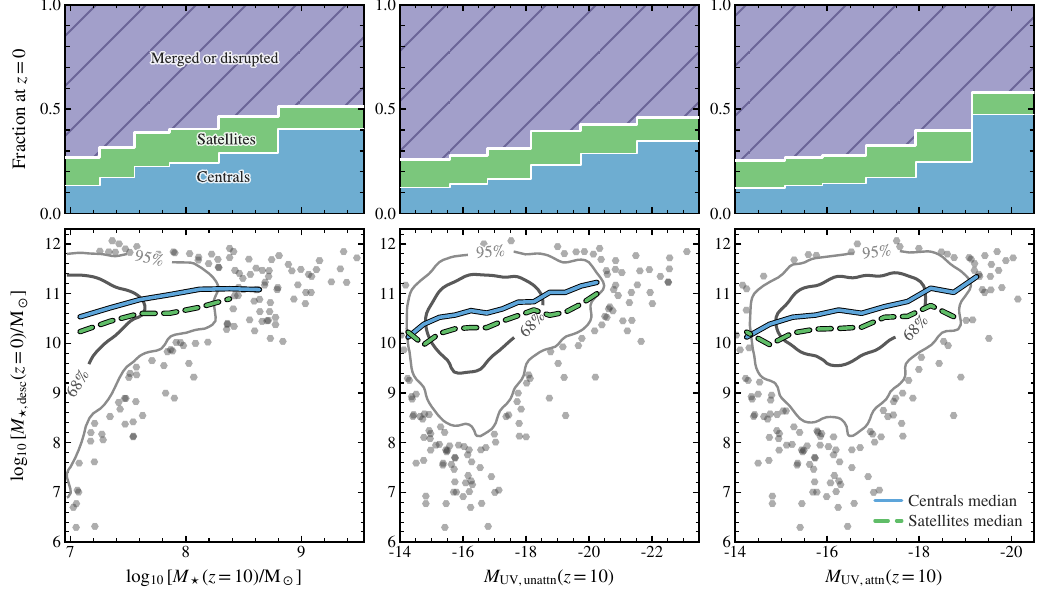}
\caption{
Connection between the properties of progenitor galaxies at $z=10$ and the stellar masses of their descendants at $z=0$ in the COLIBRE L200m6 simulation. From left to right, the progenitors are characterized by stellar mass, intrinsic (i.e. unattenuated) UV magnitude and dust-attenuated UV magnitude. The upper panels show the fraction of galaxies in each descendant fate category, using the same colours as Fig.~\ref{fig:fate_evolution}. The lower panels show the corresponding descendant stellar masses for surviving galaxies. Grey points denote individual systems, while the dark and light grey contours enclose the central 68 and 95~per~cent of the distributions, respectively. Solid and dashed curves show the median relations for surviving centrals and satellites, respectively. Although more massive or UV-brighter progenitors tend to have more massive descendants, the broad distributions demonstrate that present-day stellar mass is only weakly correlated with the stellar mass or UV luminosity of a galaxy at $z=10$.
}
\label{fig:z10vsz0}
\end{figure*}

\subsection{What do they become: stellar masses of descendants}

\begin{figure}
\includegraphics[width=\columnwidth]
{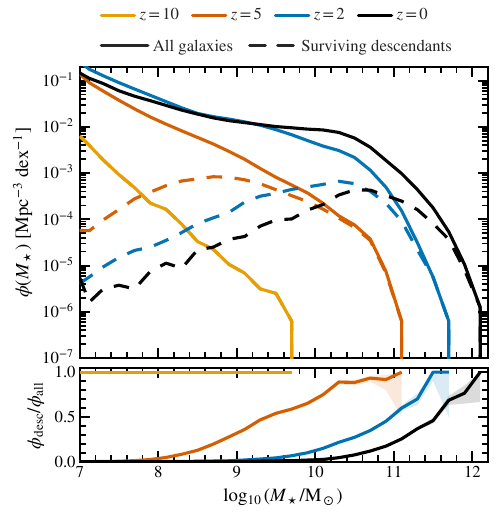}
\caption{
Contribution of galaxies selected at $z=10$ to the stellar mass function at later times. The upper panel shows the stellar mass function of all galaxies (solid curves) and of the surviving  descendants of the $z=10$ sample (dashed curves) at $z=10$, 5, 2, and 0. The lower panel shows the fractional contribution of the descendant population, $\phi_{\rm desc}/\phi_{\rm all}$, as a function of stellar mass. Here, ``contribution'' denotes a number fraction of galaxies, rather than the fraction of stellar mass in later galaxies supplied by the selected $z=10$ progenitors. Shaded regions indicate the 68 per cent Wilson binomial score confidence intervals. Galaxies already present at $z=10$ contribute mainly to the massive end of the stellar mass function at later times, while contributing little to the low-mass population.
}

\label{fig:descendant_contribution_to_smf}
\end{figure}

We now examine the present-day stellar masses associated with galaxies in the fiducial $z=10$ sample. For each surviving descendant at \(z=0\), we measure its stellar mass using bound stellar particles within a spherical aperture of radius 50 pkpc.

Fig.~\ref{fig:z10vsz0} shows the relation between three progenitor properties at $z=10$ and descendant stellar mass at $z=0$, together with the fate fractions. From left to right, the progenitors are characterized by stellar mass, intrinsic UV magnitude, and dust-attenuated UV magnitude. At fixed $M_\star(z=10)$, the surviving descendants span several orders of magnitude in $M_\star(z=0)$, while the median relation is relatively flat, with most descendants lying around $M_\star(z=0)\sim10^{10}$--$10^{11}\, {\rm M_\odot}$. The UV-luminosity panels show similarly broad distributions, although UV-brighter progenitors tend to have more massive descendants on average. Thus, all three progenitor properties provide statistical, rather than deterministic predictions of present-day stellar mass.

As shown in the lower panels, surviving centrals and satellites follow similarly broad trends, but their median descendant masses are offset. At fixed progenitor stellar mass, centrals are typically $\simeq0.5$ dex more massive at $z=0$ than satellites. This difference is consistent with galaxies that remain centrals continuing to build up stellar mass efficiently, whereas satellites may experience a reduced gas supply after infall and lose stellar mass through tidal stripping \citep{He2026COLIBREtidal}.

The upper panels show that the fate fractions vary systematically with all three progenitor properties. The fraction of surviving centrals increases from $\sim 12$~per~cent in the lowest stellar mass bin to $\sim 40$~per~cent in the highest, while the satellite fraction remains at the level of $\sim 10$~per~cent. Correspondingly, the merged/disrupted fraction decreases with progenitor mass, from $\sim 75$~per~cent to $\sim 50$~per~cent. Similar trends are found for both UV magnitudes: UV-brighter galaxies are more likely to survive to $z=0$. The total surviving fraction reaches $\sim 60$~per~cent in the brightest dust-attenuated bin. This dependence is expected in hierarchical assembly: more massive progenitors, which also tend to be UV brighter, are associated with deeper potential wells and are therefore more likely to retain a resolved descendant, whereas lower-mass systems are more easily stripped, disrupted, or incorporated into larger systems. Nevertheless, even in the highest-mass bin, about half of the galaxies selected at $z=10$ no longer survive as independent self-bound systems at $z=0$.

Having quantified how the eventual fate and present-day stellar mass of the $z=10$ galaxies depend on their initial stellar mass and UV luminosity, we next assess the contribution of their surviving descendants to the present-day galaxy population. Fig.~\ref{fig:descendant_contribution_to_smf} compares the stellar mass function of these descendants with that of all galaxies present at $z=5, 2$ and 0, and shows their fractional contribution to the stellar mass function at each selected redshift. 

The descendants occupy increasingly narrow ranges of stellar mass at decreasing redshifts. At $z=0$, there is a clear peak at $M_\star \sim 10^{10.5}$--$10^{11}\,{\rm M_\odot}$. Galaxies selected at very high redshift therefore do not evolve into the full present-day galaxy population. Instead, they are preferentially associated with intermediate- to high-mass galaxies by $z=0$.

At lower present-day masses, $M_\star \lesssim 10^{9.5}\,{\rm M_\odot}$, the contribution from the $z=10$ selected sample is negligible. This does not imply that low-mass galaxies have no high-redshift building blocks. Rather, many present-day low-mass galaxies either form most of their stars after $z=10$, or have progenitors at $z=10$ that fall below the stellar mass resolution limit of our sample. The $z=10$ population traces only a small subset of the progenitors of present-day low-mass galaxies.

At intermediate present-day masses, the contribution remains modest. Around Milky-Way mass, $M_\star(z=0)\simeq10^{10.5}\,{\rm M_\odot}$, only $7.5$ per cent of galaxies are surviving descendants of the mass-selected $z=10$ sample. When the $z=10$ sample is restricted to the UV-bright subset, this fraction falls to $0.34$ per cent. Thus, most present-day Milky Way-mass galaxies are not connected to our fiducial $z=10$ population selected here, and only a very small fraction are connected to its UV-bright subset.

At the high-mass end, the contribution from the high-redshift-selected sample becomes substantial. In particular, the ratio $\phi_{\rm desc}/\phi_{\rm all}$ approaches unity for the most massive galaxies at $z=0$, indicating that nearly all galaxies in this mass range have at least one resolved progenitor among the galaxies selected at $z=10$. In this sense, the most massive present-day systems were already represented in the early galaxy population. This does not mean, however, that the most massive galaxies at $z=0$ descend from the most massive galaxies at $z=10$. As we will show in Section~\ref{sec:backward eps}, present-day massive systems assemble from many progenitors, and their early main progenitors are generally not the most extreme objects at high redshift.

The highest-mass bins should still be interpreted with caution because of limited number statistics. In Fig.~\ref{fig:descendant_contribution_to_smf}, the shaded regions indicate the 68 per cent Wilson score confidence intervals on $\phi_{\rm desc}/\phi_{\rm all}$. These uncertainties become large where the number of galaxies is small, and individual objects can therefore noticeably affect the inferred fractional contribution. Nevertheless, the broader trend is robust: the contribution of galaxies selected at $z=10$ increases strongly with present-day stellar mass, while remaining negligible at the low-mass end.

\subsection{Where do they live: host halo masses of descendants}
\label{sec:where do they live}
\begin{figure}
\includegraphics[width=\columnwidth]
{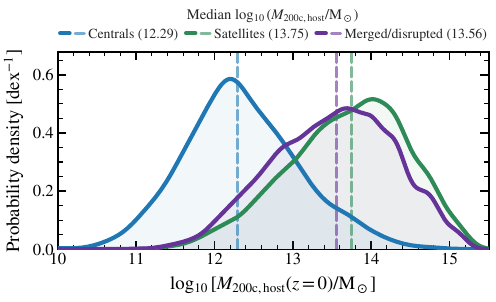}
\caption{
Distribution of $z=0$ host halo masses of galaxies selected at $z=10$, split by final fate. Curves and shaded regions show the probability density of $M_{200{\rm c},{\rm host}}$, while vertical dashed lines mark the medians, also quoted in parentheses in the legend. Surviving centrals reside in lower-mass host haloes, whereas satellites and merged/disrupted systems are typically found in galaxy groups.
}
\label{fig:z10_descendant_host_halo_mass}
\end{figure}

Fig.~\ref{fig:z10_descendant_host_halo_mass} shows the $z=0$ host halo mass distribution of both surviving and non-surviving systems, where halo mass is defined as $M_{200{\rm c}}$, the mass enclosed within a radius inside which the mean density is 200 times the critical density of the Universe. For merged or disrupted systems, we follow the HBT-HERONS descendant connection to identify their associated host halo at at $z=0$. Surviving centrals reside in haloes with median mass $\log_{10}(M_{200{\rm c}}/{\rm M_\odot}) \sim 12.3$, while surviving satellites belong to more massive host haloes, reaching $\log_{10}(M_{200{\rm c}}/{\rm M_\odot}) \sim 13.8$. Non-surviving systems, including both merged and disrupted galaxies, are concentrated in similar halo masses, with a characteristic scale of $\log_{10}(M_{200{\rm c}}/{\rm M_\odot}) \sim 13.6$.

The lower halo masses of surviving centrals reflect their distinct assembly histories. By definition, these galaxies have not been accreted into a more massive host, and therefore typically reside in relatively isolated haloes that have experienced quieter subsequent growth. In contrast, surviving satellites and non-surviving systems are associated with regions where halo growth is faster, driven by accretion and mergers that build galaxy groups. Once high-redshift galaxies enter these environments, they are more likely to become satellites and subsequently undergo tidal stripping, disruption, or merging.

The characteristic descendant host masses, $M_{200{\rm c}}\sim10^{13}$--$10^{14}\,{\rm M_\odot}$, therefore show that most of the high-redshift systems end up in group-scale haloes rather than in present-day clusters, which typically have $M_{200{\rm c}}\sim 10^{14}$--$10^{15}\,{\rm M_\odot}$. The L200m6 simulation contains 71  haloes with $M_{200{\rm c}}>10^{14}\,{\rm M_\odot}$ at $z=0$. Therefore, this preference for group-scale hosts is not due to volume limitations or  a  scarcity of cluster-mass environments.
In Section~\ref{sec:forward eps}, we show that this behaviour is well explained by EPS theory.

\subsection{Are the brightest high-redshift galaxies different? }
\begin{figure*}
\includegraphics[width=\textwidth]
{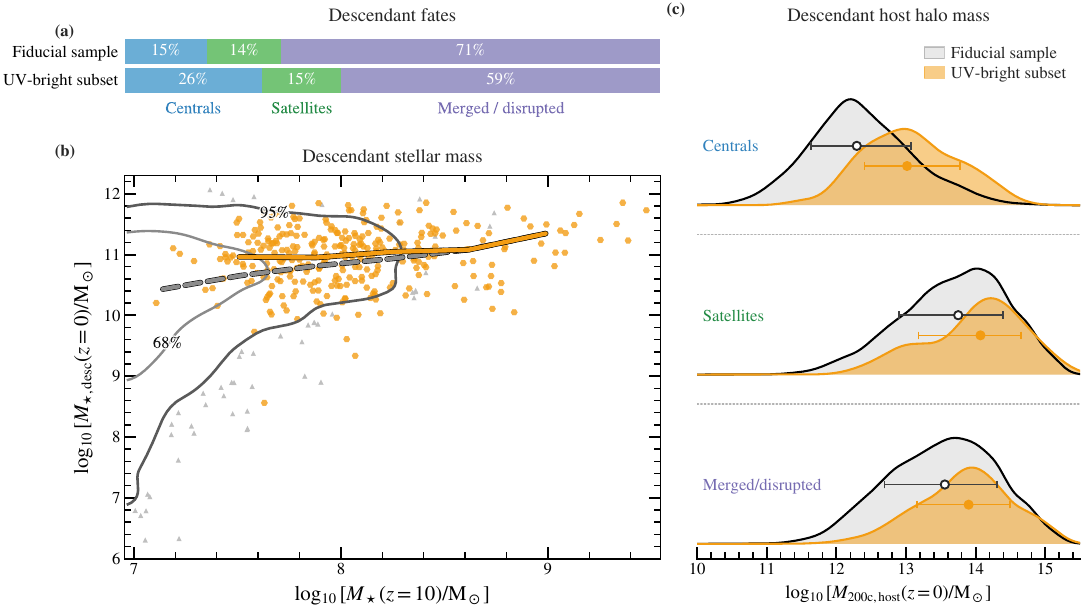}
\caption{
Comparison between the $z=0$ properties of the descendants of the fiducial $z=10$ sample and the descendants of the UV-bright subset with $M_{\rm UV,attn}<-18$. Panel (a) shows the final fate fractions of surviving centrals, surviving satellites, and merged/disrupted systems. Panel (b) shows descendant stellar mass at $z=0$ as a function of progenitor stellar mass at $z=10$ for surviving galaxies; grey triangles and contours show the fiducial sample, while orange hexagons show the UV-bright subset; dashed grey and solid orange curves show their corresponding median relations, respectively. Panel (c) compares the $z=0$ host halo mass distributions for each fate group. Black curves and open circles represent the fiducial sample, while orange curves and filled circles represent the UV-bright subset. The circles mark the medians, and the horizontal bars show the corresponding 16th--84th percentile ranges. The UV-bright subset is biased towards higher progenitor and descendant stellar masses, but its distributions overlap strongly with those of the mass-selected population.
}
\label{fig:uv_bright_vs_all_z10}
\end{figure*}

Having established the descendant statistics for the mass-selected $z=10$ population, we now ask whether UV-bright galaxies, which are more likely to be detected in high-redshift surveys, follow distinct evolutionary pathways. Fig.~\ref{fig:uv_bright_vs_all_z10} compares the mass-selected sample with the subset satisfying $M_{\rm UV,attn}<-18$. We compare three quantities: the final fate fractions, the relation between progenitor and descendant stellar mass and the $z=0$ host halo mass distributions for each fate group.

UV-bright galaxies show modest but systematic differences relative to the parent, mass-selected sample. They have a higher surviving central fraction, $26$~per~cent compared with $15$~per~cent, and a lower merged/disrupted fraction, $59$~per~cent compared with $71$~per~cent, while the satellite fraction is similar. Their descendants also have higher stellar masses on average, and reside in somewhat more massive host haloes. However, these differences are small compared with the broad scatter in the fiducial population and the two samples overlap strongly in both descendant stellar mass and host halo mass.

Thus, UV selection changes the descendant statistics rather subtly. Because $M_{\rm UV,attn}$ is correlated with stellar mass, the UV-bright cut preferentially selects more massive and actively star-forming progenitors. There is no clear evidence that UV-bright galaxies define a qualitatively distinct evolutionary pathway from the mass-selected high-redshift population.

To understand why the progenitor and descendant masses are so weakly correlated, we examined the Spearman rank correlation coefficients between various quantities linking galaxies across cosmic time. At $z=10$, several of these relations retain substantial rank information: the intrinsic UV luminosity is strongly correlated with the SFR ($\rho=0.89$), while the correlations between SFR and stellar mass and between stellar mass and host halo mass are more moderate, with $\rho=0.47$ and $0.61$, respectively. The dominant loss of rank information occurs during the subsequent evolution of the host haloes: the correlation between $M_{\rm h,host}(z=10)$ and $M_{\rm h,host}(z=0)$ is only $\rho=0.20$. The final relation between present-day host halo mass and descendant stellar mass is also broad, although a moderate correlation remains ($\rho=0.49$). These coefficients do not provide a formal decomposition of the scatter, but they indicate that the broad mapping from $M_\star(z=10)$ to $M_\star(z=0)$ is driven primarily by the weak correspondence between halo mass at $z=10$ and final halo mass, rather than by the galaxy--halo connection at either individual epoch. In the following section, we show, using the EPS framework, that this loss of rank information arises naturally from the stochastic and hierarchical growth of dark matter haloes.
\section{Interpreting descendant halo growth}
\label{sec:evolution}

\subsection{Descendant halo growth in the EPS framework}
\label{sec:forward eps}
\begin{figure*}
\includegraphics[width=\textwidth]
{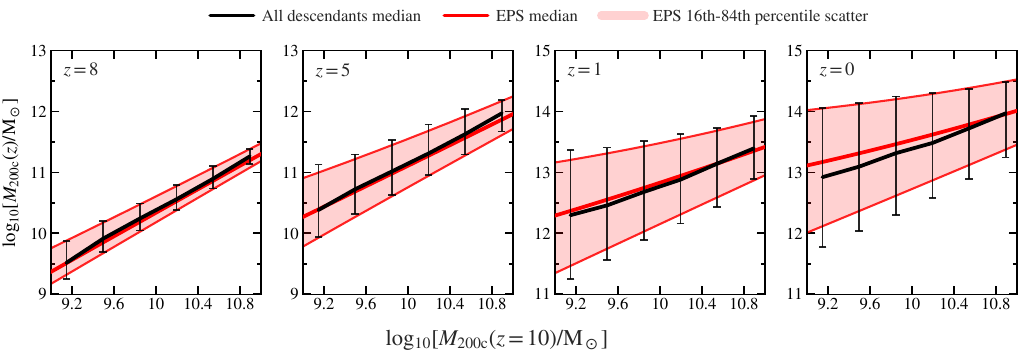}
\caption{
Evolution of descendant host halo mass as a function of progenitor halo mass at $z=10$. Each panel shows a different redshift, with the $x$-axis giving $M_{200{\rm c}}(z=10)$ and the $y$-axis giving the host halo mass of the descendants at that epoch. The black curve shows the median relation measured from the simulation, with vertical error bars indicating the 16th--84th percentile range. The red curve and shaded region show the corresponding median and 16th--84th percentile interval predicted by extended Press--Schechter (EPS) theory. The agreement between the simulation and EPS calculation shows that the median halo growth and its large scatter are natural consequences of stochastic hierarchical assembly.
}
\label{fig:forward_eps}
\end{figure*}
To understand why galaxies selected at \(z=10\) typically end up in group-scale haloes with \(M_{200{\rm c}}\sim10^{13.5}\,{\rm M_\odot}\) by \(z=0\) (Section~\ref{sec:where do they live}), we interpret the growth of their host haloes using the EPS formalism \citep{PressSchechter1974PS,Bond1991EPS,Bower1991EPS,Lacey1993EPS}. In this framework, halo assembly is described statistically as a random walk in the smoothed linear density field, expressed in terms of the variance of this field, \(S(M)=\sigma^2(M)\). A halo of mass \(M\) is associated with the first crossing of a collapse barrier, \(\omega(z)=\delta_{\rm c}/D(z)\), where \(\delta_{\rm c}\) is the critical linear overdensity for spherical collapse and \(D(z)\) is the linear growth factor of density perturbations, normalised to unity at \(z=0\). The EPS framework predicts not only the halo abundance at a given redshift, but also the conditional probability distribution connecting haloes at different epochs.

We use the forward conditional distribution of \citet[equation~2.16]{Lacey1993EPS} to compute the probability that a halo of mass, \(M_1\), at \(z_1=10\) is incorporated into a descendant halo more massive than \(M_2\) at a later redshift, \(z_2<z_1\):
\begin{equation}
P_{\rm desc}(>M_2,z_2|M_1,z_1)
=
\int_0^{S(M_2)}
f(S,\omega_2|S_1,\omega_1)\,{\rm d}S ,
\end{equation}
where \(S_1=S(M_1)\), \(\omega_i=\omega(z_i)\), and \(f(S,\omega_2|S_1,\omega_1)\) is the EPS first-crossing probability density for the descendant variance at \(z_2\), conditional on the halo mass at \(z_1\). Since \(S(M)\) decreases monotonically with halo mass, the condition \(M_{\rm desc}>M_2\) is equivalent to \(S_{\rm desc}<S(M_2)\), which sets the upper limit of the integral. We use this cumulative distribution to derive the median and percentile ranges of the expected descendant halo masses, and compare them with the simulation in Fig.~\ref{fig:forward_eps}.

The figure shows that the median descendant halo mass in the COLIBRE simulations increases steadily with cosmic time, reaching $M_{200{\rm c}}\sim10^{13}$--$10^{14}\,\mathrm{M_\odot}$ by $z=0$. At fixed initial mass, the distribution of descendant halo masses is broad, spanning more than an order of magnitude at late times. The EPS predictions (red curves) reproduce both the median trends and the large scatter seen in the simulations. This agreement indicates that the diversity of descendant halo masses arises naturally from the stochastic nature of hierarchical growth in $\Lambda$CDM.

Importantly, although descendant halo mass remains strongly correlated with progenitor mass at early times, for example at \(z=8\), this dependence weakens substantially by low redshift. The weak dependence of the \(z=0\) descendant halo mass on the progenitor mass at \(z=10\) indicates that these galaxies do not preserve a strong memory of their initial rank within the halo population. Instead, by the present day they occupy a broad range of halo masses, with a typical value of \(M_{200{\rm c}} \sim 10^{13.5}\,{\rm M_\odot}\), characteristic of galaxy groups, albeit with substantial object-to-object scatter. This behaviour provides a natural explanation for the trends discussed in Section~\ref{sec:where do they live}: most of these systems end up in galaxy groups rather than in the most massive present-day galaxy clusters.

\subsection{Extreme haloes and the loss of rank memory}
\label{sec:backward eps}

\begin{figure}
\includegraphics[width=\columnwidth]
{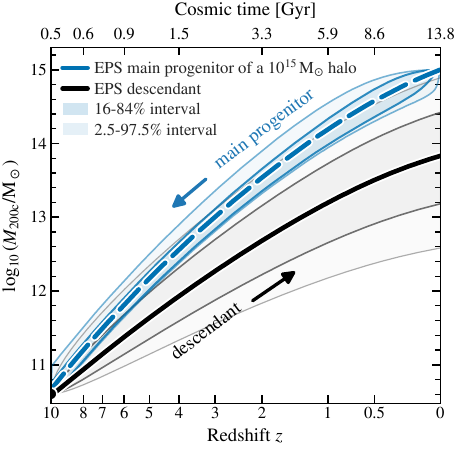}
\caption{
Halo mass growth in the extended Press--Schechter (EPS) framework. The blue curve shows the median main-progenitor history of a present-day halo with $M_{200{\rm c}}=10^{15}\,{\rm M_\odot}$, traced backwards from $z=0$ to $z=10$. The black solid curve shows the median forward descendant growth of haloes selected at $z=10$ with the same mass as this main progenitor at $z=10$. Shaded regions show the 16$^\mathrm{th}$--84$^\mathrm{th}$ and 2.5$^\mathrm{th}$--97.5$^\mathrm{th}$ percentile intervals. The two curves are not inverse mappings of one another: haloes with the same mass as the $z=10$ main progenitor of a present-day cluster typically evolve into substantially lower-mass systems by $z=0$. Thus, selecting haloes by mass at $z=10$ does not typically pick out the main progenitors of the most massive haloes today.
}
\label{fig:eps_forward_backword}
\end{figure}

Having used the forward EPS calculation to ask where haloes selected at $z=10$ are expected to end up, we now use the complementary progenitor form of the EPS formalism to address the inverse question: what were the typical high-redshift progenitors of present-day haloes? For a halo of mass, $M_0$, at $z_0=0$, the backward conditional distribution of equation~(2.15) in \citet{Lacey1993EPS} gives the mass distribution of its progenitors at an earlier redshift $z_1$. We integrate this distribution to compute the expected cumulative number of progenitors more massive than $M_1$,
\begin{equation}
N_{\rm prog}(>M_1,z_1\mid M_0,z_0=0).
\end{equation}
This quantity describes, in an ensemble-averaged sense, how many progenitor haloes above a given mass threshold are expected at redshift $z_1$ for a present-day halo of mass $M_0$.

We define a statistical most massive progenitor following \citet{GaoLiang2005MMP,Correa2015MAH,Liu2024MAH}. For a present-day halo of mass, $M_0$, we integrate the EPS progenitor distribution at redshift, $z$, from the high-mass end down to the mass, $M_{\rm th}$, at which the expected number of progenitors above this threshold is unity,
\begin{equation}
N_{\rm prog}(>M_{\rm th},z\mid M_0,z_0=0)=1 .
\end{equation}
We then estimate the most massive progenitor mass as the mean mass of progenitors in this high-mass tail,
\begin{equation}
M_{\rm MMP}(z\mid M_0) =
\frac{\int_{M_{\rm th}}^{M_0} M\,({\rm d}N/{\rm d}M)\,{\rm d}M}
{\int_{M_{\rm th}}^{M_0} ({\rm d}N/{\rm d}M)\,{\rm d}M}.
\end{equation}
This provides a statistical reference for the typical main-progenitor growth of present-day haloes, rather than the merger history of any individual object.

Fig.~\ref{fig:eps_forward_backword} illustrates the asymmetry between the backward main-progenitor history of a present-day halo and the forward descendant growth of haloes selected at $z=10$ using a cluster halo as an example. The blue curve shows the median main-progenitor history of a halo with $M_{200{\rm c}}(z=0)=10^{15}\,{\rm M_\odot}$, while the black solid curve shows the median descendant growth of haloes whose mass at $z=10$ matches that of this main progenitor. Although the two curves are anchored at the same mass at $z=10$, they diverge strongly towards lower redshift.

Forward and backward selections are therefore not inverse mappings. Haloes with the same mass as the $z=10$ main progenitor of a present-day cluster typically evolve into substantially lower-mass systems by $z=0$. Conversely, present-day cluster haloes assemble through the accumulation of many progenitors, rather than through the growth of a single, exceptionally massive high-redshift object.

This asymmetry explains the loss of rank ordering between high redshift and the present day. The most massive galaxies or haloes at $z\gtrsim 10$ should not be interpreted as straightforward progenitors of the most massive present-day clusters. Instead, stochastic hierarchical growth produces a broad mapping between early and late-time halo populations, consistent with the large scatter seen in Figs.~\ref{fig:z10_descendant_host_halo_mass}--\ref{fig:forward_eps}.

\begin{figure}
\includegraphics[width=\columnwidth]
{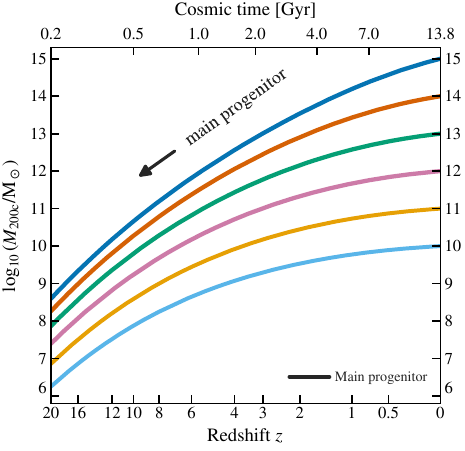}
\caption{Main-progenitor halo mass growth predicted by the EPS framework. Each curve shows the typical evolution of $M_{200{\rm c}}$ from $z=0$ to $z=20$ for haloes with present-day masses $M_{200{\rm c}}(z=0)=10^{10}$--$10^{15}\,{\rm M_\odot}$. The arrow indicates the direction followed when tracing the main progenitor backwards from the descendant halo to high redshift. The mapping between high-redshift progenitor mass and present-day halo mass is strongly compressed at $z\gtrsim10$, so early mass rank carries only limited information about the final halo rank.}
\label{fig:eps_main_progenitor_mass_grid_startz20p0}
\end{figure}

Fig.~\ref{fig:eps_main_progenitor_mass_grid_startz20p0} extends this EPS interpretation to a wider range of descendant halo masses and to higher redshift, $z=20$. Instead of considering only cluster-scale descendants, we compute the typical main-progenitor growth for haloes with present-day masses from $M_{200{\rm c}}=10^{10}$ to $10^{15}{\rm M_\odot}$. At fixed redshift, more massive present-day haloes are associated with more massive main progenitors, but the separation between the tracks is much smaller at early times than at $z=0$. The mapping between high-redshift progenitor mass and present-day halo mass is therefore compressed at $z\gtrsim10$: even haloes that become clusters by the present day had much less massive main progenitors ($M_{200{\rm c}}\sim10^{8}$--$10^{9}\,\mathrm{M_\odot}$ at $z=20$) at these redshifts. This reinforces the conclusion that selecting the most massive haloes or galaxies at very high redshift does not uniquely identify the progenitors of the most massive systems today. The final halo mass is built up through subsequent hierarchical growth, so the rank ordering of haloes at early times retains only limited predictive power of the rank ordering at $z=0$.

\section{Conclusions}
\label{sec:conclusions}

We have investigated the evolutionary fate of galaxies selected at $z \gtrsim 10$ in the COLIBRE simulations \citep{Schaye2026ColibreIntro, Chaikin2026calibration}, using \textsc{HBT-HERONS} merger trees to connect high-redshift galaxies to their present-day descendants. Our main analysis is based on the COLIBRE L200m6 simulation and focuses on galaxies selected at $z=10$ resolved by at least 5 stellar particles, $M_\star > 9\times10^{6}\,{\rm M_\odot}$, corresponding to our mass-selected sample. We also consider a UV-bright subset with $M_{\rm UV,attn}<-18$, where the dust-attenuated rest-frame UV magnitudes are computed by applying the SKIRT radiative transfer code to the COLIBRE outputs. Our aim has been to ask what present-day population is probed when galaxies are observed at very early cosmic times, particularly in the context of \textit{JWST} detections of UV-bright systems at $z \gtrsim 10$. Our main conclusions are as follows.

\begin{itemize}

\item The dust-attenuated UV magnitude is correlated with stellar mass, but with substantial scatter. UV selection is therefore mass-biased, but not equivalent to a simple stellar mass selection (Fig.~\ref{fig:sample_selection_z10}).

\item Most galaxies selected at $z=10$ do not survive as distinct self-bound systems to $z=0$. The surviving fraction decreases steadily with cosmic time, while the merged/disrupted component grows and becomes the dominant outcome by the present day. By $z=0$, only about one third of the original population remains as resolved descendants, split equally between centrals and satellites (Fig.~\ref{fig:fate_evolution}).

\item The mapping from $z=10$ progenitor properties to present-day stellar mass is broad: at fixed stellar mass or UV magnitude, descendants span several orders of magnitude, although most reach $M_\star(z=0)\sim10^{10}$--$10^{11}\,{\rm M_\odot}$. For the UV-bright subset, the median descendant stellar mass is $\sim10^{11}\,{\rm M_\odot}$, nearly independent of progenitor stellar mass at $z=10$, despite substantial scatter for individual galaxies. These progenitor properties are therefore only weak predictors of descendant stellar mass (Fig.~\ref{fig:z10vsz0}). This diversity is also illustrated by the mock visualisations, where galaxies with similar high-redshift masses can evolve into very different present-day systems (Fig.~\ref{fig:mock_images}).

\item Galaxies already present in the resolved $z=10$ sample contribute mainly to the massive end of the later stellar mass function. Their contribution is negligible at low present-day stellar masses, but increases strongly towards high masses, indicating that the resolved high-redshift population traces an important part of the assembly history of present-day massive galaxies rather than of the full $z=0$ galaxy population (Fig.~\ref{fig:descendant_contribution_to_smf}).

\item The descendants typically reside in present-day galaxy groups. Surviving centrals are found in relatively lower-mass haloes, whereas satellites and merged/disrupted systems preferentially occupy more massive hosts. For the full descendant population, the characteristic host mass is $M_{200{\rm c}}\sim10^{13.5}\,{\rm M_\odot}$, corresponding to group-scale rather than massive cluster-scale environments (Fig.~\ref{fig:z10_descendant_host_halo_mass}).

\item UV-bright galaxies have modestly higher survival fractions and somewhat more massive descendants than the mass-selected population. However, their fate fractions, descendant stellar masses, and host halo masses overlap strongly with those of the parent sample. UV selection therefore changes the descendant statistics quantitatively, mainly by selecting more massive and actively star-forming progenitors, but it does not define a qualitatively distinct evolutionary pathway (Fig.~\ref{fig:uv_bright_vs_all_z10}).

\item The evolution of descendant host halo masses is consistent with the expectations from EPS theory. The EPS calculation reproduces both the median halo growth and the large scatter seen in the simulation, supporting the interpretation that the high-redshift to low-redshift connection is intrinsically statistical (Fig.~\ref{fig:forward_eps}).

\item Forward and backward halo selections are not inverse mappings of one another. For example, while the main progenitor of a $z=0$ cluster is typically a massive $z=10$ halo, the descendants of these haloes are not typically clusters. Therefore the most extreme galaxies or haloes at $z\gtrsim10$ should not be interpreted as straightforward progenitors of the most massive present-day clusters (Fig.~\ref{fig:eps_forward_backword}).

\end{itemize}

Our results extend earlier efforts to connect very high-redshift systems with the local Universe. \citet{ChenYangyao2023JWSTDesc} used the constrained dark-matter simulation ELUCID to connect massive galaxies at $z\simeq 8$ to their descendant haloes and to the ancient stellar populations expected in present-day cluster centrals. \citet{Lushengdong2025GALFORM} used the semi-analytic galaxy formation model \textsc{galform} to trace UV-bright galaxies selected at $z=14$, finding descendants with a broad range of stellar masses, typically in group-scale haloes and often as satellites by the present day. On the whole, their conclusions are similar to ours.

Our analysis provides a complementary hydrodynamical view of this descendant connection. In COLIBRE, galaxy growth, metal enrichment, feedback, the multiphase interstellar medium, and dust evolution are followed within a single galaxy-formation model, while dust-attenuated UV magnitudes are computed with SKIRT radiative transfer post-processing. This allows us to compare a stellar mass-selected sample with a UV-bright subset selected using dust-attenuated magnitudes, and to quantify the galaxy-level fate of each system as a surviving central, surviving satellite, or merged/disrupted descendant. The UV-bright progenitors are biased towards more massive and more actively star-forming galaxies at high redshift, but their present-day outcomes strongly overlap with those of the parent sample. The broad descendant distribution is therefore not simply an artefact of selecting galaxies by stellar mass, nor is it significantly modified by applying an observationally motivated UV selection.

A second contribution of this work is the explicit EPS interpretation of the descendant halo growth. The EPS calculation reproduces both the median growth and the large object-to-object scatter of the descendant host haloes, showing that the weak rank ordering between $z\gtrsim 10$ progenitors and their present-day descendants is an expected consequence of stochastic hierarchical assembly. Taken together, these results indicate that galaxies selected at $z\gtrsim10$, including UV-bright systems accessible to \textit{JWST}, should be interpreted statistically rather than as one-to-one progenitors of a single present-day galaxy class. Their descendants span a broad range of stellar masses, environments, and survival histories, reflecting the hierarchical build-up of intermediate- to high-mass galaxies.
\section*{Acknowledgements}

AP was supported by funding from the European Research Council (ERC) under the European Union’s Horizon 2020 research and innovation programmes (grant agreement no. 818085 GMGalaxies). EC acknowledges support from the Science and Technology Facilities Council (STFC) through consolidated grant ST/X001075/1. KAO acknowledges support from the Royal Society through a Dorothy Hodgkin Fellowship (DHF/R1/231105). FH acknowledges funding from the Netherlands Organization for Scientific Research (NWO) through research programme Athena 184.034.002. ABL acknowledges support by the Italian Ministry for Universities (MUR) program “Dipartimenti di Eccellenza 2023-2027” within the Centro Bicocca di Cosmologia Quantitativa (BiCoQ), and support by UNIMIB’s Fondo Di Ateneo Quota Competitiva (project 2024-ATEQC-0050). The authors declare no conflicts of interest.

We thank Anna Durrant for providing the SKIRT scripts and for helpful explanations of the COLIBRE–SKIRT pipeline. This work used the DiRAC@Durham facility managed by the Institute for Computational Cosmology on behalf of the STFC DiRAC HPC Facility (\url{https://www.dirac.ac.uk}). The equipment was funded by BEIS capital funding via STFC capital grants ST/K00042X/1, ST/P002293/1, ST/R002371/1, and ST/S002502/1, Durham University, and STFC operations grant ST/R000832/1. DiRAC is part of the UK National e-Infrastructure.


\section*{Data Availability}

The public version of the SWIFT code is available at \url{https://www.swiftsim.com}. The SWIFT modules related to the COLIBRE galaxy formation model will be integrated into the public version following the public release of COLIBRE. The HBT-HERONS halo finder is available at \url{https://HBT-HERONS.strw.leidenuniv.nl/}. The \textsc{soap} catalogue pipeline is publicly available at \url{https://github.com/SWIFTSIM/SOAP}.



\bibliographystyle{mnras}
\bibliography{example} 




\appendix
\section{Convergence of fate fractions}
\label{sec:appendix fateconvergence}
\begin{figure*}
	\includegraphics[width=\textwidth]
    {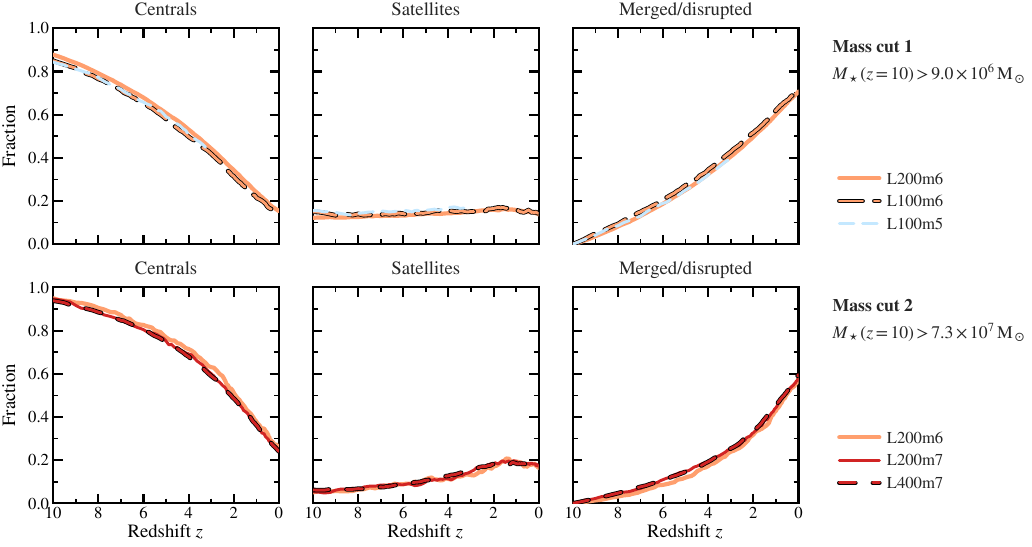}
    \caption{
    Resolution and volume convergence of the descendant fate fractions for galaxies selected at $z=10$ using two stellar mass thresholds. The upper row shows galaxies selected with $M_\star(z=10)>9.0\times10^6\,{\rm M_\odot}$, corresponding to $5\,m_{\rm gas,m6}$, while the lower row shows galaxies selected with $M_\star(z=10)>7.3\times10^7\,{\rm M_\odot}$, corresponding to $5\,m_{\rm gas,m7}$.  For each mass cut, panels show the fraction of the original $z=10$ sample classified as surviving centrals, surviving satellites, or merged/disrupted systems as a function of redshift. Lines show different COLIBRE runs. The L100m5 simulation has not yet been evolved to $z=0$, and is therefore absent at $z<3$. The trends are consistent across simulations: the central fraction decreases towards low redshift, the satellite fraction remains subdominant, and the merged/disrupted fraction grows strongly by $z=0$.}
    \label{fig:fate_fraction_convergence_5particles}
\end{figure*}

To assess the robustness of the descendant fate fractions measured in our fiducial L200m6 run, we repeat the analysis using matched stellar mass cuts across simulations with different resolutions and volumes. This test addresses three possible sources of bias: artificial disruption, in which galaxies classified as disrupted at lower resolution may survive as resolved self-bound remnants at higher resolution; resolution-dependent galaxy properties, which may affect their subsequent fates even in the absence of artificial disruption; and finite-volume variation, in which different box sizes sample different large-scale environments. It therefore tests whether the fate fractions used in the main analysis are robust against numerical resolution and volume effects.

Figure~\ref{fig:fate_fraction_convergence_5particles} shows the evolution of the fate fractions for galaxies selected at $z=10$ using two stellar mass thresholds. The upper row adopts $M_\star(z=10)>9.0\times10^6\,{\rm M_\odot}$, corresponding to $5\,m_{\rm gas,m6}$, and compares L200m6 with the smaller-volume L100m6 and higher-resolution L100m5 simulations. The lower row adopts $M_\star(z=10)>7.3\times10^7\,{\rm M_\odot}$, corresponding to $5\,m_{\rm gas,m7}$, and compares L200m6 with the lower-resolution L200m7 and larger-volume L400m7 simulations.

The broad behaviour is consistent across simulations. For both mass cuts, the central fraction decreases steadily towards low redshift, the satellite fraction remains subdominant but increases mildly, and the merged/disrupted fraction grows strongly towards $z=0$. The fate fractions typically differ by less than 5~per~cent between runs, both for the higher threshold and for the lower, 5-particle threshold. This is consistent with the independent convergence study of \citet{He2026COLIBREtidal}, who analysed the tidal evolution of satellite galaxies in COLIBRE and found that stellar tidal tracks are reasonably well converged between the m6 and m5 resolutions for $m_{\star,\rm peak}>10^7\,{\rm M_\odot}$, close to our fiducial mass threshold. We therefore conclude that the main descendant fate fractions are not dominated by resolution or finite-volume effects, supporting the use of the fiducial L200m6 selection adopted in the main text.

\FloatBarrier
\section{Results for Descendants of \lowercase{$z=12$} galaxies}

\label{sec:appendix z12 results}
\begin{figure}
\includegraphics[width=\columnwidth]{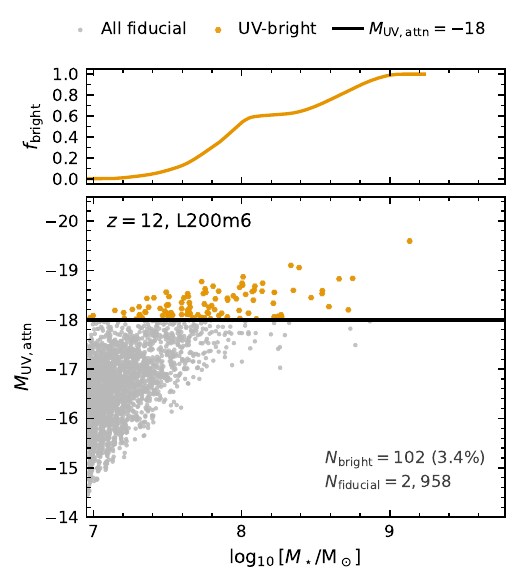}
\caption{
Same as Fig.~\ref{fig:sample_selection_z10}, but for galaxies selected at $z=12$. At $z=12$, only 102 of the 2,958 mass-selected galaxies ($3.4$~per~cent) are UV-bright, with the UV-bright subset preferentially occupying higher stellar masses.}
    \label{fig:sample_selection_z12}
\end{figure}

In the main text we focus primarily on galaxies selected at $z=10$, where the sample size is large enough to provide robust statistics. Here we repeat the same descendant analysis for the resolved $z=12$ sample in the L200m6 simulation. This extends the analysis to earlier cosmic times, although the smaller sample size means that the resulting trends are noisier.

For the $z=12$ analysis, we use the same five-star-particle selection criterion, corresponding to $M_\star > 9\times10^6\,{\rm M_\odot}$ in L200m6. This yields 2958 resolved galaxies, of which 102 satisfy our UV-bright threshold, $M_{\rm UV,attn}<-18$ (Fig.~\ref{fig:sample_selection_z12}).

\begin{figure}
	\includegraphics[width=\columnwidth]{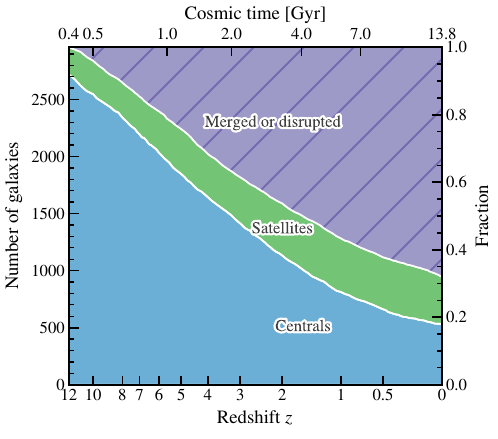}
    \caption{
    Same as Fig.~\ref{fig:fate_evolution}, but for galaxies selected at $z=12$. Thus, the $z=12$ sample follows the same qualitative fate evolution as the $z=10$ sample, with non-survival by $z=0$ again dominating despite larger statistical uncertainty.}
    \label{fig:fate_evolution_z12_selection}
\end{figure}
Fig.~\ref{fig:fate_evolution_z12_selection} shows the evolution of the fate fractions for galaxies selected at $z=12$. The qualitative behaviour is the same as for the $z=10$ sample: the fraction of systems that remain as surviving centrals decreases steadily with time, the satellite fraction remains subdominant, and the merged/disrupted fraction grows towards low redshift. By $z=0$, only a minority of the original $z=12$ population survives as distinct self-bound galaxies, while most systems have either  merged into a more massive system or become disrupted. Thus, for galaxies selected at $z=12$, the dominant outcome is incorporation into larger structures rather than survival as isolated present-day galaxies.

\begin{figure*}
	\includegraphics[width=\textwidth]
    {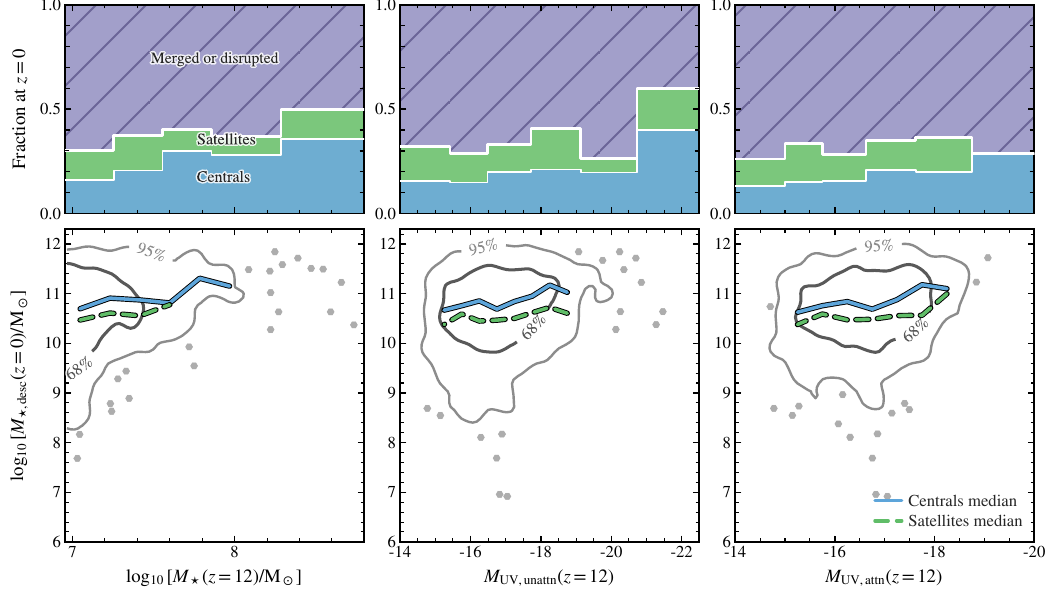}
    \caption{
    Same as Fig.~\ref{fig:z10vsz0}, but for galaxies selected at $z=12$. This confirms that the weak and highly scattered mapping between early properties and present-day descendant mass is already present for galaxies selected at $z=12$.}
    \label{fig:Mstar_z12vsz0}
\end{figure*}
Fig.~\ref{fig:Mstar_z12vsz0} shows the relation between three progenitor properties at $z=12$ and the stellar mass of the descendant galaxy at $z=0$. As in the $z=10$ case, the mapping is broad: at fixed $M_\star(z=12)$, descendants span several orders of magnitude in $M_\star(z=0)$. Surviving centrals tend to occupy the upper part of the descendant stellar mass distribution, while satellites typically have lower present-day stellar masses, consistent with reduced late-time growth after accretion into larger haloes. Despite the smaller sample and narrower stellar mass range at $z=12$, the main conclusion is unchanged: present-day stellar masses retain only weak memory of the stellar masses of their very high-redshift progenitors.

\begin{figure}
	\includegraphics[width=\columnwidth]
    {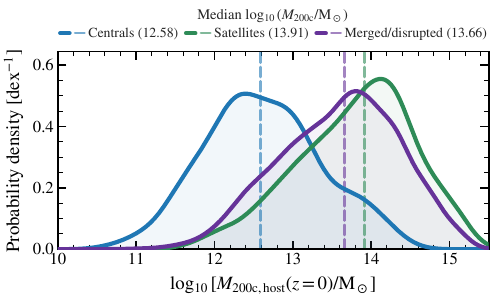}
    \caption{
    Same as Fig.~\ref{fig:z10_descendant_host_halo_mass}, but for galaxies selected at $z=12$. The $z=12$ descendants occupy similar present-day environments to the $z=10$ sample, with satellites and non-surviving systems preferentially ending up in group-scale haloes.}
    \label{fig:z12_descendant_host_halo_mass}
\end{figure}
Fig.~\ref{fig:z12_descendant_host_halo_mass} shows the $z=0$ host halo mass distributions for galaxies selected at $z=12$. Surviving centrals reside in relatively lower-mass haloes, with a median $\log_{10}(M_{200{\rm c}}/{\rm M_\odot}) \simeq 12.6$. In contrast, surviving satellites are found in more massive environments, with a median $\log_{10}(M_{200{\rm c}}/{\rm M_\odot}) \simeq 13.9$. The merged/disrupted systems occupy similar galaxy groups, with a median $\log_{10}(M_{200{\rm c}}/{\rm M_\odot}) \simeq 13.7$. Thus, as for the $z=10$ sample, satellites and non-surviving systems are mainly associated with group-scale haloes by the present day, while surviving centrals occupy systematically lower-mass hosts.

Overall, the $z=12$ results reinforce the conclusions drawn from the larger $z=10$ sample. Most galaxies selected at very early times do not survive as distinct present-day systems, and those that do survive span a broad range of stellar masses and environments. Although the $z=12$ sample is smaller and therefore noisier, it supports the same physical picture: the connection between galaxies at $z\gtrsim10$ and their present-day descendants is statistical and highly stochastic, rather than one-to-one.


\FloatBarrier
\label{lastpage}
\end{document}